# Regulating for an AI Commons

## Adopting, sustaining, and defending people-first innovation trajectories

**Abhineet Nayyar[a], Mallory Knodel[b], Kimberlee Weatherall[c], and Gunn Jiravuttipong[d]**

*Abstract — Technological principles of openness, interoperability, and decentralization are integral to building an AI economy that is accountable to the public and truly responsive to local needs and constraints.*

*However, Big Tech's firm control over essential elements of the stack (such as chips, cloud, models, data, and talent) limits valuable alternatives from reaching the scale necessary for sustained growth. Worse still, hyperscalers continue to enclose and extract value from publicly funded and volunteer-driven technologies without adequately distributing these returns. We see consolidated power within a few entities and 'enshittification' of once-transformative innovations. Embedding ethical micro-level tools (like open data repositories or small-language models) into workflows is promising, but without a consistent innovation strategy, they also often end up perpetuating and being appropriated by the same exploitative trajectories.*

*Instead, this whitepaper outlines a three-pillared holistic and intentional policy approach that i) actively pushes for the adoption of cooperative technologies in existing and emerging systems, ii) sets up collective institutional structures to sustain and maintain them in the long-term, and iii) introduces regulatory proposals that limit private entities from disempowering them through uncontrolled enclosure and extraction. Through desk research and purpose-specific case studies, we argue for the importance of such an approach in the Global South, where structural constraints around infrastructure, financing, and policy governance make it difficult for AI commons and cooperative AI tools to emerge and survive.*

*The paper concludes with key open questions for campaigners, policy advocates, researchers, and grassroots organizers to consider and explore while working towards peoples' solidarity within and between Global South countries.*

[a] Senior Associate, Research & Policy Engagement, IT for Change (https://itforchange.net/)
[b] Cryptographer and Executive Director, Social Web Foundation (https://socialwebfoundation.org/)
[c] Co-Director, Centre for AI, Trust, and Governance, and Professor of Law, University of Sydney (https://www.sydney.edu.au/)
[d] Research Fellow, Thailand Development Research Institute (https://tdri.or.th/en/)

[a] Senior Associate, Research & Policy Engagement, IT for Change (https://itforchange.net/)
[b] Cryptographer and Executive Director, Social Web Foundation (https://socialwebfoundation.org/)
[c] Co-Director, Centre for AI, Trust, and Governance, and Professor of Law, University of Sydney (https://www.sydney.edu.au/)
[d] Research Fellow, Thailand Development Research Institute (https://tdri.or.th/en/)

# I. Framing the problem

*Market concentration within a handful of Big Tech Transnational Companies (TNCs), combined with a growing weaponization of global policy instruments, has homogenized the arc of AI innovation into serving the interests of private capital and one-way value extraction.*

From the browser wars of the late 1990s, to the subsequent era of online walled gardens controlled by a few private entities, the last thirty years have seen multiple waves of decline, fragmentation, and enclosure of the digital commons.[1] [2] That the web of today looks starkly different from its early vision of an open, interoperable, and federated information ecosystem has long been a matter of concern for technologists and researchers alike.[3] The highly extractive and asset-light business model of data profiling which fuels the modern commercial internet, combined with market traits like network effects, economies of scale, and algorithmic management, has consistently allowed ‘first-movers’ to dominate a range of digital markets,[4] [5] and in some cases leverage that dominance into new areas, including AI.

Emerging AI paradigms could accelerate these trends: we are seeing various attempts at enclosure and concerted commercial efforts to build dominance up and down the AI stack. But AI also raises novel questions for the rights of those it exploits during development, use and expansion. In a geopolitical moment that is incentivising strategic competitiveness rather than mutual collaboration, Big Tech hyperscalers are actively seeking out more lucrative opportunities to entrench dominance at the different levels of hardware and digital infrastructures on which our digital lives are built.[6]

Be it the compute, the cloud, or the models, a handful of corporations have come to use business tactics like bundling and vertical integration, financial strategies like circular investing and creative accounting, and lobbying efforts directed at the deregulation of labour, environmental, and taxation standards to accelerate their own growth and exert control over the entire stack.[7] [8] As developing countries and middle powers seek to join the AI race, they have been faced with a challenging choice in determining where to concentrate investment and/or regulation to enable local innovation and economic activity — compete with Big Tech entities (which, given the potential costs, may not be possible for most countries), partner with them to use their hardware and other services (including models) (and risk dependence and potential lock-in), or build everything (perhaps at some less advanced level) to side-step the largest systems.[9]

---

[1] Stalder, F. “The Decline of the Digital Commons.” May 2026. Available here.
[2] Halpin, H. “The Hidden History of the Like Button: From Decentralized Data to Semantic Enclosure.” August 2023. Available here.
[3] Doctorow, C. “Microincentives and Enshittification.” July 2023. Available here.
[4] Henrijke, R., Adriaans, I., Klinge, T.J., et al. “The Big Techification of Everything”. October 2021. Available here.
[5] Tan, JS and Kathleen, T. “Cloud Capitalism and the AI Transition.” December 2025. Available here.
[6] Khanal, S., Zhang, H., and Taiehagh, A. “Why and how is the power of Big Tech increasing in the policy process? The case of generative AI.” January 2025. Available here.
[7] Sleiman, L.M. and Harrigan, K.R. “Strategic Flexibility for Hyperscalers: Lessons from the AI Infrastructure Build-Out.” March 2026. Available here.
[8] Lancieri, F., Edelson, L., and Bechtold, S.. “AI Regulation: Competition, Arbitrage & Regulatory Capture.” July 2025. Available here.
[9] Huang, S. “Securing Agency and Managing Trade-offs in the Age of AI: Strategic Choices for Asia-Pacific Middle Powers.” January 2026. Available here
[a] Senior Associate, Research & Policy Engagement, IT for Change (https://itforchange.net/)
[b] Cryptographer and Executive Director, Social Web Foundation (https://socialwebfoundation.org/)
[c] Co-Director, Centre for AI, Trust, and Governance, and Professor of Law, University of Sydney (https://www.sydney.edu.au/)
[d] Research Fellow, Thailand Development Research Institute (https://tdri.or.th/en/)

Recent improvements in the performance of open weights models may change the picture, potentially bringing lower costs for users of AI, though still not addressing other dependencies such as the need for access to compute.

Ongoing experience suggests that none of these approaches may provide real autonomy to the rest of the world: given the vast resources involved, for most countries direct competition remains almost impossible. App-layer or model-layer integration, on the other hand, risks value transfer from, and further exploitation of, local actors. Demands for full-stack localization rooted in national sovereignty, which are proposed as a way to escape Big Tech dependence, also provide little solace to countries with low industrial maturity and no control over the necessary intellectual capital.[10] Even if it was feasible, on its own, it would only prevent foreign capture of the AI stack. Countries attempting the home-grown approach still face growing energy demands, market consolidation, data extractivism, and worker exploitation.

More importantly, this lack of autonomy, or as it has been framed in Australia, a lack of *agency*, is not just a strategic risk for nation-state sovereignty.[11] Experience from network, platform, and utility (NPU) studies has repeatedly shown that dependence for essential needs on a handful of corporations, especially those you cannot effectively penalize, leads to an under-regulated ethical and safety landscape.[12] [13] By paying insufficient attention to the profit motive and the consolidating objectives that influence technical decisions made by digital firms, and failing to imagine alternative designs, policy regimes in developing countries also end up perpetuating false binaries between innovation and regulation.

Addressing this interplay of market pressures, technical standards, design choices, and fundamental rights requires a re-imagination of the options and regulatory choices that steer technology towards an arc that puts people first. What we need is a shift from monolithic, market-driven extractive impulses, to a pluralistic conception of AI commons that can enable the distinct needs people actually have.

## II. Fostering a people's AI stack

*Although commons-based design practices, and alternative models of ownership and control have gained prominence, a lack of regulatory and institutional support has hindered their ability to compete with, and more importantly, displace the Big Tech's hegemony.*

[10] Switaj, C. and Ping, J. "Artificial Intelligence and Geopolitics: Middle Powers in the Age of Algorithmic Power." December 2025. Available here.
[11] Tech Policy Design Institute. "Expanding AI Sovereignty to AI Agency." June 2026. Available here.
[12] Narechania, T.N. and Sitaraman, G. "An Antimonopoly Approach to Governing Artificial Intelligence." January 2024. Available here.
[13] Ricks, M., Sitaraman, G., Welton, S., et al. "Networks, Platforms, and Utilities: Law and Policy." 2022. Available here.
[a] Senior Associate, Research & Policy Engagement, IT for Change (https://itforchange.net/)
[b] Cryptographer and Executive Director, Social Web Foundation (https://socialwebfoundation.org/)
[c] Co-Director, Centre for AI, Trust, and Governance, and Professor of Law, University of Sydney (https://www.sydney.edu.au/)
[d] Research Fellow, Thailand Development Research Institute (https://tdri.or.th/en/)

In her landmark book, *Re-Enchanting the World: Feminism and the Politics of the Commons*, Silvia Federici highlights how the commons — including the "new forms of social cooperation being produced [...] on the internet" — are vital for survival against global capitalist extraction.[14]

If today's AI tools are, indeed, to serve the needs of all the people of the world, we would need ways to break from the excessively profit-driven incentive structure driving much of the capital flows into the stack. This is particularly true for the expensive compute and cloud infrastructure layers, which are currently being operated by a small number of corporations and seek to maximize market returns and not use values.[15]

One promising alternative is to be found in the many forms of collective ownership emergent in the modern internet economy. These include worker and social cooperatives gaining rapid prominence in the platform and data industries, as well as instances of public control over critical digital infrastructures. But while shared forms of management and governance provide key foundations on which to build the vision of a commons, practicalizing and sustaining them in the throes of market liberalization requires the necessary technical and legal interventions.

## Technical preconditions and the limits of techno-neutrality

From the early years of the internet, which saw the likes of the HTTP and SMTP protocols, to the more recent embrace of digital public goods and infrastructures, open, interoperable, and federated architectures have been integral to building shared and equitable technological systems. These architectures have acted as potential buttresses against commons enclosure, as evidenced by a range of state- and community-driven projects in the recent past.

Brazil's **Pix** system, for example, is a large-scale payments system run on interoperable protocols to prevent direct market capture by any single entity. Pix is operated by Brazil's national bank, thereby allowing for more public access to and scrutiny over the technology's specifications. Similarly, open protocols such as **ActivityPub** and **ATProto** enable cross-platform communication, reducing switching costs and weakening network effects that would otherwise lock users with dominant social media platforms. In a third example, "**Public AI**" seeks to overcome economic constraints (like rising business costs) and ethical concerns (like undisclosed data scraping) through high-performing open source models and community-first licensing.[16] [17]

The ability of such cumulative and cooperative systems to strike the difficult balance between innovation and regulation, makes them engines of growth, but also enduring targets of exploitation and monopoly enclosures, which surfaces the limit of a singular techno-neutral approach.[18]

Unfortunately, the reconceptualisation of openness under the Enterprise AI moment's profit-first mandate has recast economic and social values of technology, and enabled it to be weaponised for extraction and capture of digital ecosystems.[19] Contrary to the neutrality and interoperability principles it originated in, rhetorical and strategic claims on behalf of "**openness**" have been leveraged to entrench platform monopolies and the same walled gardens we find ourselves locked into today.[20]

[14] Federici, S. "Re-Enchanting the World: Feminism and the Politics of the Commons." 2018. Available here.
[15] Fisher, A., and Streinz, T. "Confronting Data Inequality." May 2022. Available here.
[16] Tarkowski, A. "NOODL — An experiment in equitable data licensing." April 2026. Available here.
[17] Mozilla. "Open Source AI 2026." July 2026. Available here.
[18] Knodel, M. "From the Commons to Rentierism: Governance Against a Closed Internet." April 2026. Available here.

[a] Senior Associate, Research & Policy Engagement, IT for Change (https://itforchange.net/)
[b] Cryptographer and Executive Director, Social Web Foundation (https://socialwebfoundation.org/)
[c] Co-Director, Centre for AI, Trust, and Governance, and Professor of Law, University of Sydney (https://www.sydney.edu.au/)
[d] Research Fellow, Thailand Development Research Institute (https://tdri.or.th/en/)

Physical and digital infrastructures, from submarine cables and operating systems, to cloud services, certificate authorities, DNS resolvers, and dominant social graphs have become sites where control over essential services yields a handful of firms ongoing rent with minimal productive contribution.[21] [22] And it is concentration of resources at these upstream sites that have become the foundation of today's AI ecosystems, where control over data, compute, discovery, and distribution significantly amplifies the same extractive and harmful dynamics.

## Coalescing around cross-border governance objectives and standards

Policy interventions like investigation and enforcement by domestic competition regulators and sovereignty mandates in AI strategies have emerged most prominently in many debates as possible responses to market enclosures. But their success towards building a shared economic landscape has remained minimal due to enforcement challenges, the delays inherent in these processes, and the market power entrenched by dominant enterprises.[23] [24] At the same time, supranational internet governance bodies, such as the IETF or the IEEE, continue to represent multistakeholderism in a polarized global order, making them marginal solutions for addressing market concentration, at best, and at worst, captured legal arrangements.

Finally, and most importantly, the high-stakes geo-eco-political race underpinning the AI paradigm has left emerging economies with a policy approach that lacks a complementary understanding of governance measures and technical architectures, particularly in the context of the US-China dynamic.[25] Even the most open, interoperable, and federated alternatives — from RISC-V chip architectures and OS models, to inter-cloud portability standards — are far from being adopted consistently, let alone adopted sufficiently broadly to present a sustainable alternative to the enclosing tendencies of the hyperscalers and the Big Tech.

The current policy landscape is diverse, with promising individual elements, but highly disintegrated. It spans technical standards, cooperative ownership structures, and regulatory measures. Enforcement happens in isolation, if at all. In the following chapter we propose an alternative that ties these instruments together in a holistic governance approach for a shared AI commons. Unlike goals of national sovereignty that force the people of the world to assimilate into, or compete with, the extractive and enclosing global AI paradigm, we envision a future where AI stacks consist of shared and accessible infrastructures hosting public interest and rights-respecting goods for the collective benefit.

---

[19] Kilic, B., & Knodel, M. "Big Tech Redefined the Open Internet to Serve Its Own Interests." July 2025. Available here; Widder, D., Whittaker, M., West, S.. "Open (for business): Big Tech, concentrated power, and the political economy of open AI". August 2023. Available here.
[20] Berk, G. & Saxenian, A. "Architectures of Participation." 2022. Available here.
[21] Knodel, M. "From the Commons to Rentierism: Governance Against a Closed Internet." April 2026. Available here.
[22] Saito, K., & Sasaki, R. "Rentier capitalism, technofascism and the destruction of the common." October 2025. Available here.
[23] Winant, G. "Marxism and Antitrust: A Provocation." September 2025. Available here.
[24] Cohen, Julie E. "Between truth and power." 2019.
[25] Martina, M. "US to tell partners they must pick sides in AI race with China". August 2026. Available here.
[a] Senior Associate, Research & Policy Engagement, IT for Change (https://itforchange.net/)
[b] Cryptographer and Executive Director, Social Web Foundation (https://socialwebfoundation.org/)
[c] Co-Director, Centre for AI, Trust, and Governance, and Professor of Law, University of Sydney (https://www.sydney.edu.au/)
[d] Research Fellow, Thailand Development Research Institute (https://tdri.or.th/en/)

# III. Discussing the three pillars

*With this context, we put forth three key institutional objectives for an AI commons: adoption, sustenance, and defense. To that end, we include indicative policy responses beneath each, such that they hold critical potential for the survival of an AI economy rooted in principles of mutual care and solidarity.*

Our proposal rests on three primary pillars, each signifying a key strategic objective in popularizing and normalizing a commons-centered AI ecosystem. The objective of this chapter is to introduce these pillars, establish their significance for our discussion, and articulate key policy interventions that could be implemented to meet their ends.

The idea of the subsequent discussion is not as much to present a comprehensive review of all policy interventions that could support an AI commons movement, but more to conceptualize and support, via specific case studies, a framework that can further enable such experiments and possible coordination/collaborations between states. The examples listed below also provide proofs of concept for policy measures that regulators, public-interest technologists, and governments across the world can borrow from and re-imagine within their local contexts.

## **Pillar 1 | Adoption** — Using cooperative technologies in existing and emerging systems

The entrenched market dominance of a handful of companies across the AI stack also comes with a reinforcing tendency. The first movers' advantage and the network effects inherent to digital and platform markets—combined with an investment regime that rewards centralized scale over adaptive cooperativism—creates a fertile ground for monopolization. In this context, it is critical that the mere existence of open, interoperable, and decentralized alternatives not be conflated with collaborative digital ecosystems. Instead, we must imagine concrete steps to customize, securitize, and diversify the AI commons while also enabling their integration into real-time workflows: all with a view to promoting the **adoption** of open technologies that is critical to achieving the Sustainable Development Goals.[26]

We present below two proposals to that end, each accompanied with learnings from particular case studies.

- **Explicit preference for cooperative tools, technologies, and ownership structures in public procurement contracts**
  As more and more state functions and services get enmeshed with processes of digitisation, the reliance of public authorities on privately owned and maintained digital tools and infrastructures increases manifold.[27] This dependence poses a threat to national security, state sovereignty, and cyber-resilience. More generally, such partnerships allow hyperscalers and Big Tech entities to capture explicit, and implicit, information about the country's citizens, and the state's experience of providing essential services. Control over such foundational knowledge enables these firms to expand further across the stack, and beyond limited contracts, further narrowing opportunities for possible competitors.

[26] See p8-9, UN Secretary-General's Roadmap for Digital Cooperation. 2020. Available here.
[27] Bhattacharya, A. "Pushing back from Big Tech: Africa's hard road to AI sovereignty." May 2026. Available here.
[a] Senior Associate, Research & Policy Engagement, IT for Change (https://itforchange.net/)
[b] Cryptographer and Executive Director, Social Web Foundation (https://socialwebfoundation.org/)
[c] Co-Director, Centre for AI, Trust, and Governance, and Professor of Law, University of Sydney (https://www.sydney.edu.au/)
[d] Research Fellow, Thailand Development Research Institute (https://tdri.or.th/en/)

In contrast, the adoption of free, open-source, interoperable and publicly maintained systems could allow public authorities to escape such patterns of potential lock-in and unilateral extraction. Limited experiences in countries like India, Brazil, and Kenya, indicate possibilities for governments interested in reducing their reliance on Big Tech intermediaries.[28] Creating an AI commons, though, would require going beyond one-off solutions built on certain cooperative methods and public monies, to making this a core part of the public procurement process.

From climate commitments to ethical supply chains and market competition, public procurement has historically played a critical role in redirecting innovation trajectories, and could be deployed to promote cooperative technologies, as well.[29] This would mean embedding technical principles of openness and interoperability, and public-interest requirements of accountability and collaborative design, into public contracts and requests-for-proposals (RfPs), as well as upskilling public servants to build confidence in technology procurement and carefully assessing - and where possible, amending or removing - other procurement requirements that tend to discourage government entities from adopting open technologies. In addition to building institutional autonomy, this move provides a clear demand signal to nudge established players towards more collaborative tool development.

Tanzania's multi-year push for replacing proprietary systems with open-source alternatives within public departments is a significant example on this front. As the Minister of Information, Communication and Technology of the Republic of Tanzania, put it recently, 90% of their government systems are built on open-source tools.[30] India's policy experience also provides a key precedent, especially the state of Kerala, which has specifically pushed for providing clear preference to the adoption of Free and Open Source Software (FOSS) in all government projects.[31]

This pathway can, and should, be further expanded to cover more cooperative principles appropriate to different levels of the AI stack and to different countries' circumstances, including but not limited to decentralised decision-making architectures, public auditability, and shared ownership structures, especially for the underlying/upstream infrastructure that can otherwise enable concentration and monopolisation.

- **Socialized, accountable, and end-to-end support for digital public goods and infrastructures**

---

[28] Scaria, A G, Mishra, S., et al. "The Rise of FOSS in India: Empirical Evidence and Insights from Cross-Sectoral Case Studies." July 2025. Available here.
[29] Morris, D. and Veiberg, C B "Public procurement as a tool to address human rights risks in the use of digital technology to deliver essential public services." February 2024. Available here.
[30] United Nations Transcripts. "Open Source Program Offices for Good - UN Open Source Week 2026 (Part 5): Testimony at 17:46." June 2026. Available here.
[31] Scaria, A G, Mishra, S., et al. "The Rise of FOSS in India: Empirical Evidence and Insights from Cross-Sectoral Case Studies." July 2025. Available here.

[a] Senior Associate, Research & Policy Engagement, IT for Change (https://itforchange.net/)
[b] Cryptographer and Executive Director, Social Web Foundation (https://socialwebfoundation.org/)
[c] Co-Director, Centre for AI, Trust, and Governance, and Professor of Law, University of Sydney (https://www.sydney.edu.au/)
[d] Research Fellow, Thailand Development Research Institute (https://tdri.or.th/en/)

In spite of the mainstreaming of digital public goods (DPGs) and digital public infrastructures (DPIs) in policy discourse, their adoption, even within individual countries, lacks consistent application of commons-centred principles.[32]

As existing research has noted, these products and protocols are not neutral artefacts, but instead reflect explicit techno-legal choices that are shaped by, and in turn, shape, the political economies of their respective ecosystems. The challenges posed by this dynamic are, possibly, most evident in India's approach to DPGs and DPIs — which exist largely to solve for economic and market alignment challenges.[33] As a result of such narrow mandates, most of them focus on interoperability as a key feature, but often disregard aspects like source code openness, public ownership and auditing.

But the global landscape of DPGs and DPIs contains a diverse range of unique organizational structures. Brazil's experience with Pix (its state-owned payments infrastructure) and more recently, with the Cadastro Ambiental Rural (CAR) (its rural-environmental data registry) provide two examples of alternative DPIs. CAR, in particular, is an interesting state-led effort to build land, property, and environment databases, for agri-business opportunities, as well as research and policy enforcement.[34] [35] Built and maintained by public IT firms in the country, some of the DPI's modules have been further open-sourced into DPGs that other nations can use.[36]

This approach to shared value generation is critical to the adoption of cooperative technologies. While preferential procurement measures affect the state's adoption of FOSS tools, a collaborative and accountable network of public institutions that drive research, development, and innovation in DPGs&Is is critical to building a bottom-up AI commons independent of market pressures and demands. At the same time, regional leaders across the Global South must also align together and undertake bold steps towards endowing the compute layer into a more distributed ownership structure.

---

[32] DPGs and DPIs are distinct, but complementary concepts —

- 'Digital public goods' here refers to free, interoperable, and open source tools, open data systems, public AI stacks, and and open content collections that are available for anyone to use, copy, and modify (see, for example, the UN Secretary General's Roadmap for Digital Cooperation, June 2020, available here).
- 'Digital public infrastructures' refer to a set of shared digital 'rails' that are secure, non-excludable, interoperable, often standardized and publicly governed, which enable a range of state activities — such as inclusive delivery of government and market services at scale or enforcing regulatory mandates, among others. While operational experience varies, DPIs often combine open technology standards with accountable governance frameworks to encourage market competition, service delivery, or shared control of critical systems to track resource consumption and value distribution (see, for example, definitions put forth by the Centre for Digital Public Infrastructure (here), the UNDP (here), and the Open Future Foundation (here)).

[33] IT for Change. "Recovering the 'Public' in India's Digital Public Infrastructure Strategy: Policy Brief." January 2025. Available here.

[34] de Godoy, C M et al. "From Brazil to the world: Open digital infrastructure for climate cooperation." March 2026. Available here.

[35] Samdub, M T. "Brazil's Cadastro Ambiental Rural and the Vision for Progressive Digital Public Infrastructure." 2026. (Forthcoming)

[36] Government of Brazil. "Rural Environmental Registry Registration Module." November 2025. Available here.

[a] Senior Associate, Research & Policy Engagement, IT for Change (https://itforchange.net/)

[b] Cryptographer and Executive Director, Social Web Foundation (https://socialwebfoundation.org/)

[c] Co-Director, Centre for AI, Trust, and Governance, and Professor of Law, University of Sydney (https://www.sydney.edu.au/)

[d] Research Fellow, Thailand Development Research Institute (https://tdri.or.th/en/)

## **Pillar 2 | Sustenance** — Building collective institutional structures to ensure long-term maintenance of alternatives

A core aspect of the existing AI economy is its deep entanglement with a data-maximalist business model and a finance-driven form of neoliberalism, both of which further the extractive technological tendencies and sustain ongoing AI development.[37] An AI stack rooted in principles of shared ownership and value distribution would also require robust policy apparati to protect, and fund its ongoing development maintenance, and sustain its goals of public-interest innovation. Furthermore, the mechanisms and interventions listed under this pillar would also play an important role in (partially) liberating the commons from volatile market forces, whose profit-first framing has consistently enclosed open and collective knowledge structures.

Below, we introduce two examples of policymaking and institutional governance that could be considered indicative of this pillar.

- **Global taxation and disbursal of profit accrued from digital workstreams and affiliated activities**
  In September 1998, the General Council of the World Trade Organization adopted the work programme on electronic commerce, which led to the establishment of the moratorium on customs duties on electronic transmissions.[38] In 1998, the digital economy was at a very early stage of its development: this predates most social media, and mainstream development of AI: policies adopted at that time may make less sense now. Nevertheless, since then, WTO member states have largely refrained from imposing cross-border taxes on “electronic transmissions,” including data flows, online payments, cloud computing, and other digital services.

  Despite (or perhaps because of) its long-standing existence, however, the moratorium has come under increasing scrutiny from Global South countries whose shares of tax revenues from the digital economy remain sparse. One estimate suggests that between 2014 and 2025, the effective corporate tax rate of the ‘Silicon Six’ was about 19%, significantly lower than the global average of 27%.[39]

  Attention is often trained on the unprofitability of elements of the AI stack (and the potential for an AI bubble). But unprofitability at one level can coincide with concentration and bottlenecks at other layers of the stack - and consequent profit extraction, now or in the future. The ‘AI race’ remains undeniably important in entrenching the dominance of cloud hyperscalers like Google, Microsoft, Amazon, Meta, Alibaba, Huawei, and Oracle.[40]

  To sustain a knowledge and AI commons in this moment requires leveraging Big Tech’s renewed interest in developing economies for the common good. To that end, digital taxation standards and measures (cross-border and otherwise) should be considered as potentially important policy moves by middle powers, and other regionally cooperative countries, to redirect some of the firms’ profits towards more shared purposes.

---

[37] Venkatasubramanian, S. “How AI and surveillance capitalism are undermining democracy.” August 2025. Available here.
[38] World Trade Organization. “Work Programme on E-Commerce, Moratorium.” Available here.
[39] FairTax Foundation. “The Silicon Six and their enduring global tax gap.” April 2025. Available here.
[40] Chandran, R. “The Great AI Divide: Navigating U.S. and Chinese dominance”. June 2026. Available here.

[a] Senior Associate, Research & Policy Engagement, IT for Change (https://itforchange.net/)
[b] Cryptographer and Executive Director, Social Web Foundation (https://socialwebfoundation.org/)
[c] Co-Director, Centre for AI, Trust, and Governance, and Professor of Law, University of Sydney (https://www.sydney.edu.au/)
[d] Research Fellow, Thailand Development Research Institute (https://tdri.or.th/en/)

Countries like Indonesia, Colombia, Kenya, and Turkey, have already attempted a range of unilateral measures — in the form of Digital Services Tax (DSTs), equalization levies, Significant Economic Presence taxes, or withholding taxes, among others.[41] While most have been met with an irate US, the presence of these efforts indicates a growing support for profit-sharing methods, creating a fertile ground for potential collective action.[42] [43] Recent moves by US states like Virginia and North Carolina to remove tax exemptions and/or impose taxes on data centres, are also worth noting.[44] [45]

In addition to supporting local and sustainable alternatives, these revenues could also contribute to long-term capacity building within the open source community and the buildout of key cloud infrastructure and R&D capabilities.

- **Subsidized and reliable provisioning of AI-ready compute for commons-centred architectures and public-interest use-cases**
  An unfortunate outcome of the ongoing AI race and its reliance on compute-heavy automation has been the growing cost of accessing processing and memory units.

  Short depreciation cycles of GPUs, combined with complicated and patent-protected manufacturing processes, currently make it difficult to produce affordable compute at scale, forcing countries to rely on NVIDIA, AMD, Intel, and a handful of chip manufacturing firms. While hyperscalers like Google, Amazon, and Microsoft also produce some processing units, their hosting and deployment of this compute is tied to their respective cloud environments.

  In order to sustain alternatives including open AI , there are a number of options. For example, countries that procure and run compute at scale could subsidize their usage for specific applications and formats that yield them long-term advantages and build the innovation commons: that is, compute subsidies and access could be conditional.[46] An alternative currently being discussed in Australia is tying permissions for building data centres with conditions like the reservation of some compute for Australian, public, and/or research uses.[47]

  One prominent instantiation of public GPU provisioning can be seen in China's 'National Unified Computing Power Network' — a government initiative that aims to build an internet-like architecture of compute resources for users across the country.[48] Reports suggest that the latest node of this network, which came online this year, hosts over 100,000 computing cards for commercial users.[49] Similarly, the IndiaAI mission, launched by the Indian government in 2025, aims to provide over 45,000+ GPUs to local startups and educational institutions at affordable rates.[50] Europe's aspirations to build a CERN-like distributed computing network for research institutions is yet another indication of public provisioning of processing power.[51]

---

[41] FairTax Foundation. "The Silicon Six and their enduring global tax gap." April 2025. Available here.
[42] Office of the United States Trade Representative. "Section 301 Investigation Report on Turkey's Digital Services Tax." January 2021. Available here.
[43] Office of the United States Trade Representative. "2026 National Trade Estimate Report on Foreign Trade Barriers". March 2026. Available here.
[44] The Virginia legislature recently enacted a budget creating a new Data Center Electricity Consumption Tax: H.B. 30, Laws, 2026, at § 3-5.24.
[45] S.B. 257, Laws 2026, § 44.4(a) amending N.C. GEN. STAT. § 105-164.13(55)
[46] Sitaraman, G. and Parek, K. "The Global Rise of Public AI." May 2025. Available here.
[47] Andrew Charlton, ANU School of Public Policy Annual Lecture. August 2026. Available here.
[48] Center for Security and Emerging Technology. "Action Plan for the High-Quality Development of Computing Power Infrastructure." November 2023. Available here.
[a] Senior Associate, Research & Policy Engagement, IT for Change (https://itforchange.net/)
[b] Cryptographer and Executive Director, Social Web Foundation (https://socialwebfoundation.org/)
[c] Co-Director, Centre for AI, Trust, and Governance, and Professor of Law, University of Sydney (https://www.sydney.edu.au/)
[d] Research Fellow, Thailand Development Research Institute (https://tdri.or.th/en/)

While these examples largely remain purpose-agnostic, it is not impossible to imagine a portion of such rationalized compute being used for the sustenance of the AI commons — for example, by subsidizing access for open-source and interoperable alternatives, or by providing technical support to cooperative-first ownership schemas and rights-respecting use-cases.

## **Pillar 3 | Defence** — Resisting private sector extraction and enclosure of the AI Commons

The final pillar of a policy environment for supporting commons-based alternatives to the enclosed, proprietary AI stack is **defence.**

There is a critical need to learn from the challenges of the Free and Open Source Software (FOSS) movement, and recognise the ways in which profit-driven innovation introduces negative incentives and externalities that threaten the sustenance of the knowledge commons. On one hand, the ease of using modular code and data repositories reduces entry barriers, making commons critical for market innovation. But a lack of just and adequate compensatory mechanisms for using these (purportedly) public goods, and for the costs imposed by such use, can leave them reliant on private capital and render their technical neutrality obsolete.

For instance, the Android mobile operating system claims to be a 'FOSS' technology that should technically allow multiple companies to develop competing builds.[52] But Google's acquisition of the project, and its search market dominance, has enabled Google to bundle its services with the pre-installed Android version. And since most mobile users would be unable and, possibly, unwilling to replace Google entirely, the 'open' nature of the system is overruled, or at the very least, significantly undermined by its incorporation into and distribution via a market economy.[53] We list this case because it points us towards two policy areas that we discuss below under this pillar.

- ***Ex-ante* and stricter enforcement of digital market competition and antitrust law on an international level**
  The platform economy of the previous decade posed two main challenges to principles of competition regulation. First, the direct and indirect network effects that come with controlling a mediating software like a social media service or an online marketplace provided an advantage to capable first movers. Second, the economies of scale that accompany the maintenance of a digital infrastructure allowed these first movers to grow rapidly and outpace conventional methods of antitrust enforcement.[54]

[49] Xuanmin, L. "China's national supercomputing internet 'core node' goes online, offering over 100,000 AI computing cards." July 2026. Available here.
[50] IndiaAI. "Table for the number of empanelled GPU providers, across NVIDIA, AMD, Intel, AWS, and Google." Available here.
[51] Confederation of Laboratories for Artificial Intelligence Research in Europe. "CERN for AI." Available here.
[52] Android Open Source Project. "Content license." Available here.
[53] Competition Commission of India. "CCI imposes a monetary penalty of Rs. 1337.76 crore on Google for anti-competitive practices in relation to Android mobile devices." October 2022. Available here.

[a] Senior Associate, Research & Policy Engagement, IT for Change (https://itforchange.net/)
[b] Cryptographer and Executive Director, Social Web Foundation (https://socialwebfoundation.org/)
[c] Co-Director, Centre for AI, Trust, and Governance, and Professor of Law, University of Sydney (https://www.sydney.edu.au/)
[d] Research Fellow, Thailand Development Research Institute (https://tdri.or.th/en/)

The regulatory response to these features has been a critical development across the Global North and across a few sparse pockets of the Global South.[55] From development of nuanced theories of harm to the shift towards *ex-ante* enforcement, all the way to explicit recognition of Big Tech entities as 'monopolies', it is clear that the policy outlook towards digital market regulation seems to be shifting.[56]

However, it is essential that these learnings are extended to the AI stack, because many platform oligopolies still hold a dominant position within it.[57] Google, Microsoft, and Amazon are not just market leaders in their respective platform markets, but also provide most of the cloud infrastructure for training and inference of AI models. The resulting concentrated control over resources necessary for extending the knowledge commons poses a key challenge in the form of future lock-ins, extractive pricing practices, and self-preferential deployment.

Within this context, examples from developing countries like India's draft Digital Competition Bill (DCB) represent an important precedent. Although yet to be ratified by the Parliament, the DCB aims to include cloud services under its scope. If enacted, it would allow the state to demand *ex-ante* review of hyperscalers that dominate the cloud market.[58]

To remedy de-commonification, normal de-privatization is not always sufficient, and preventing it in the first place would be a critical intervention for the growth of DPGs and DPIs. But crucially, we should not limit our activity under this pillar to the prevention of future monopolization. It is equally important to pursue the "commonizing" or opening/ensuring of access to critical oligopoly assets, such as geospatial data, a global neutral search index, or patents held by a handful of technology firms.

- **Creation of commons-centred knowledge sharing and usage mechanisms to prevent extractive enclosures**
  In *The Rulers: Corporate Power in the Age of AI and the Cloud*, Cecilia Rikap studies the networked capitalism of the digital economy, and the ways that ownership of intangible property, such as intellectual property, facilitates concentration and the formation of monopolies. The book's study of the intangible capital held by the Big Tech (the data, the software, the intellectual property) as key revenue sources is an integral framing for the AI commons movement, for two reasons.[59]

[54] Narechania, T.N. and Sitaraman, G. "An Antimonopoly Approach to Governing Artificial Intelligence." January 2024. Available here.
[55] Jiravuttipong, G. "The Global Race to Rein in Big Tech". May 2026. Available here.
[56] Organisation for Economic Co-operation and Development. "Theories of Harm for Digital Mergers – Note by Austria to the DFEA Competition Committee." May 2023. Available here.
[57] European Commission. "Commission launches market investigations on cloud computing services under the Digital Markets Act." November 2025. Available here.
[58] Ministry of Corporate Affairs, Government of India. "Report of the Committee on Digital Competition Law." February 2024. Available here.
[59] Christiaens, T. "Big Tech and the rise of intellectual monopolies." August 2026. Available here.
[a] Senior Associate, Research & Policy Engagement, IT for Change (https://itforchange.net/)
[b] Cryptographer and Executive Director, Social Web Foundation (https://socialwebfoundation.org/)
[c] Co-Director, Centre for AI, Trust, and Governance, and Professor of Law, University of Sydney (https://www.sydney.edu.au/)
[d] Research Fellow, Thailand Development Research Institute (https://tdri.or.th/en/)

One, the AI supply chain is dominated by firms maintaining their market power through enclosing the knowledge commons, including training datasets and algorithms, chip designs, programming tools and libraries, and cloud software platforms. They may partner with and contract out to manufacturing and construction entities with technical know-how, but their stronghold is largely due to the property rights that accompany their intangible capital.

Two, much like competition, IP law has also failed to prevent value concentration and monopolisation in the current AI moment. Consider the many copyright violations claimed by creative and knowledge workers, and digital archives that have been turned down in the favor of AI companies using their work for model training and fine-tuning. Though this clearly points to the limits of existing IP law in securing collective knowledge claims, it should also be unsurprising because copyright law has always aimed to protect the market for one's labour, not the labour itself.[60]

An AI commons movement, though, must address issues of labour, markets, and property, alike, and the current IP regime's gaps present an opportunity to explore courageous solutions. For example, the copyleft concept within the FOSS movement aims to maintain the openness of works built on public and open-source artefacts, which is useful in resisting knowledge enclosures within the infrastructure layers, too. Alternative licensing regimes that support cooperativism, like NOODL and the CC Signals, combined with social labour funds that sustain open and creative work also indicate potential policy pathways for the application and model layers.[61] [62]

## IV. Concluding remarks and open questions

The ongoing commodification of open-source and open-weights AI models indicates the market's convergence towards a potential business strategy that may make AI diffusion a cost-effective strategy for enterprises. However, open and, often free, models, by themselves, also remain insufficient in resisting the market power of a handful of countries and their Big Tech hyperscalers.

First, the unequal distribution of skills between individuals and countries, means that the cloud computing companies that currently own or invest heavily in frontier AI labs, are likely to stay essential providers of infrastructure and platforms to run models for popular use. As a result, a commodified model layer would reinforce the market concentration in the layer beneath it. Secondly, without adequate institutionalization, the FOSS software movement continues to remain vulnerable to volatile market forces and extractive interests of dominant private entities. More importantly, the shift towards open models does not necessarily mean a shift towards public-interest use, and it is essential that the AI commons not be used to percolate the same harms of algorithmic profiling, discrimination, surveillance, and weaponization we see abound with proprietary tools.

---

[60] Agarwal, A. "Beyond Copyright: Why Intellectual Property regimes cannot address AI's displacement of creative labor." June 2026. Available here.
[61] Tarkowski, A. "NOODLE: An experiment in equitable data licensing." April 2026. Available here.
[62] Creative Commons. "AI and the Commons." Available here.

[a] Senior Associate, Research & Policy Engagement, IT for Change (https://itforchange.net/)
[b] Cryptographer and Executive Director, Social Web Foundation (https://socialwebfoundation.org/)
[c] Co-Director, Centre for AI, Trust, and Governance, and Professor of Law, University of Sydney (https://www.sydney.edu.au/)
[d] Research Fellow, Thailand Development Research Institute (https://tdri.or.th/en/)

To that end, we list below a few open questions for campaigners, policy advocates, researchers, and grassroots organizers to consider and explore, so that we can together build a strategy for realizing this cooperativist technological future.

1. How reliable is the concept of national sovereignty towards nudging countries to reconsider their relationships with the Big Tech and hyperscalers? Is there a better narrative that resists polarization between workers and communities of different countries, who play a variety of roles in sustaining the AI economy?
2. While this paper focuses on knowledge enclosures in the digital domain, the intellectual property rights and free trade rules that empower them are broader concepts with histories and conflicts in other sectors, too. What do we need to tie our observations and proposals to the larger requirements of Global South countries?
3. As we've seen over the last decade, DPGs and DPIs have come to be defined in a variety of ways, including some that essentially convert the state into a data collector and service provider. How does our vision of the AI commons prevent such extraction and harmful use, particularly against marginalized genders, ethnicities, castes, and groups?
4. How can we reimagine regenerative innovation trajectories and leverage tech choices with attention to planetary boundaries and equitable development considerations?

As the "AI race" forces the global order to shift from its decades-old unipolar hierarchy, it becomes increasingly important for the rest of the world to re-imagine its digital future. However, the task in front of us is not an easy one. The rifts within and between the Global South world make it difficult for interests and expectations to align with that of an AI commons, and it is, consequently, important for us to discuss and identify actionable pathways that can bring into existence this paper's proposals.

[a] Senior Associate, Research & Policy Engagement, IT for Change (https://itforchange.net/)
[b] Cryptographer and Executive Director, Social Web Foundation (https://socialwebfoundation.org/)
[c] Co-Director, Centre for AI, Trust, and Governance, and Professor of Law, University of Sydney (https://www.sydney.edu.au/)
[d] Research Fellow, Thailand Development Research Institute (https://tdri.or.th/en/)